%% file: main.tex
\documentclass{article}
\usepackage{graphicx} 
\usepackage{amsmath}
\usepackage[margin=1in]{geometry}
\usepackage{lineno}
\usepackage{xcolor}
\usepackage{mathtools}
\graphicspath{{../}}

\usepackage[
  backend=bibtex,
  style=nature,
  uniquelist=false,
  minbibnames=2,
  minnames=2,
  maxnames=6,
  natbib=true,
]{biblatex}
\title{Midlatitude cloud radiative effects reinforce storm tracks on seasonal timescales}
\author{%
Or Hadas$^{1,*}$
}

\date{}

\begin{document}

\maketitle

\begin{center}
\small
$^{1}$Department of Earth and Planetary Sciences, Weizmann Institute of Science, Rehovot 7610001, Israel\\[0.5em]
$^{*}$Corresponding author. Email: or.hadas@weizmann.ac.il
\end{center}
\begin{abstract}
\input{Abstract}
\end{abstract}
\input{Introduction}
\input{Results}

\input{Conclusions}

\input{Methods}
\input{Acknowledgment}

\printbibliography
\newpage
\input{Supplementary}
\end{document}

%% file: Abstract.tex
Midlatitude storm tracks play a central role in Earth's climate by transporting heat from low to high latitudes. Their strength is therefore set by the meridional temperature gradients, traditionally linked to differential solar heating. This framework predicts a substantial seasonal reduction in storm activity, as meridional insolation gradients vanish during summer. Yet in the Southern Hemisphere, storm activity decreases by only about 40\% from its seasonal maximum. Here, we show that cloud radiative effects contribute to the seasonal maintenance of storm activity by reinforcing the temperature gradients that sustain it. Satellite observations reveal that sunlight reflected by midlatitude clouds in early summer creates a substantial meridional gradient in surface heating, despite the nearly uniform summer insolation. Idealized aquaplanet simulations then show that shortwave cloud radiative effects reinforce meridional sea-surface temperature gradients, thereby strengthening storm activity primarily during late summer and autumn, while longwave cloud effects partly offset this response. To interpret these results, we develop a simple theoretical model linking storms, clouds, and sea-surface temperature gradients. The model captures the seasonal response and demonstrates the importance of two emergent cloud properties: the maximum attainable cloud albedo and the sensitivity of cloud cover to storm activity. Together, these findings indicate that cloud radiative feedbacks help maintain the thermal gradients that sustain storm activity on seasonal timescales. More broadly, they reveal a strong coupling among storms, clouds, and the ocean spanning distinct spatial and temporal scales.

%% file: Introduction.tex
\section*{Introduction}

Midlatitude storm tracks are a central component of Earth’s climate system, transporting heat and moisture poleward and shaping the planetary radiation budget through the extensive cloud fields they generate \cite{Peixoto1992,stephens2005cloud,Hartmann2015global,loeb2009toward,tselioudis2020midlatitude,ceppi2015connections}. Their strength depends on the meridional temperature gradients that drive baroclinic growth \cite{charney1947,Eady1949}, which are ultimately maintained by meridional gradients in solar heating \cite{Stone1978,hotta2011significance,Hartmann2015global}. 

\begin{figure}[t]
    \centering
    \includegraphics[width=0.5\linewidth]{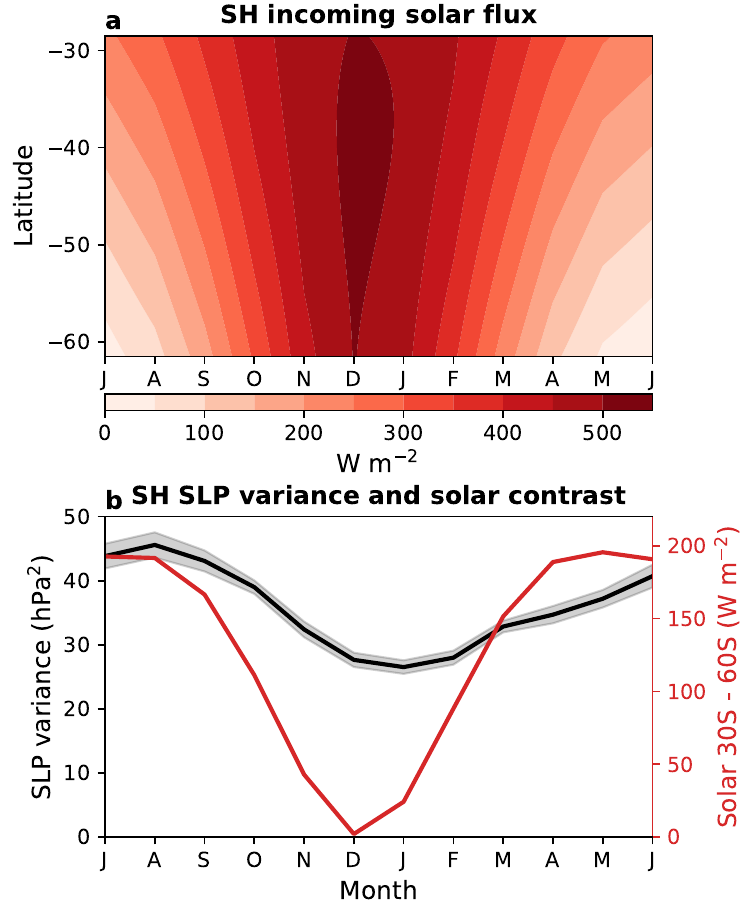}
    \caption{\textbf{Weak summertime insolation gradients contrast with persistent SH storm activity} (a) Incoming solar radiation as a function of month and latitude in the SH midlatitudes. (b) The insolation difference between latitudes 30-60$^\circ$S (red) and the sea-level pressure (SLP) variance (black), averaged monthly, zonally, and between latitudes 30–60$^\circ$S in the SH, shown as a function of month. Insolation data are from CERES, while SLP data are from ERA-5. Shading indicates 95\% confidence interval (CI, Methods).}
    \label{fig:1}
\end{figure}

This balance is especially delicate in summer, when incoming solar radiation becomes nearly uniform across the midlatitudes, producing only weak radiative forcing of meridional temperature gradients (Fig.~\ref{fig:1}a,b red). Yet storm activity does not vanish: Southern Hemisphere (SH) storm-track activity decreases by only about 40\% from its winter maximum, even as the meridional insolation gradient nearly disappears (Fig.~\ref{fig:1}b, black and red). This mismatch between weak summertime insolation gradients and persistent storm activity raises a fundamental question: what maintains SH summer storm tracks when the direct radiative forcing of midlatitude temperature gradients is weak? Previous studies have shown that the weak seasonal cycle of the SH storm track is consistent with the seasonal energy budget and is reinforced by the meridional gradient of sea surface temperature (SST) \cite{shaw2018moist}, which can be maintained, for example, by the large heat capacity of the ocean mixed layer \cite{barpanda2020surface}.

Embedded within the storm tracks themselves is a potentially important source of radiative forcing. Individual storms generate extensive cloud fields through coherent ascent, frontal lifting, and post-frontal subsidence \cite{lau1995satellite,tong2025predicting,Field2007}. Although they are typically viewed as a consequence of storm activity \cite{Grise2016understanding,Hadas2023,blanco2023cloud}, their large radiative impact \cite{Tselioudis2006,Grise2019,joos2019warm} raises the possibility that they also contribute to maintaining the circulation that generates them.

Previous research on cloud radiative effects (CRE) on the midlatitude circulation in the present climate has shown that atmospheric CRE modifies storm thermal structure and baroclinic growth on synoptic timescales, with effects depending on cloud vertical structure \cite{keshtgar2022cloud,voigt2023tug}, and shortens southern annular mode persistence on intraseasonal timescales \cite{vishny2024impact,liu2025vertically}. In the annual mean, atmospheric CRE strengthens storm activity and affects jet position, with substantial contributions from tropical clouds \cite{li2015influence,watt2017local}. Surface CRE is less understood, although SW CRE biases are linked to biases in SH jet and storm-track positions \citep{ceppi2012southern,priestley2023drivers}. Clouds also shape the circulation response to climate change \cite{voigt2015circulation,ceppi2016clouds,albern2020role}. However, previous studies have not isolated the seasonal cloud–storm feedback while combining atmospheric and surface CRE, interactive cloud–radiation coupling, and a focus on midlatitude clouds.

Here, we address this problem using a hierarchy of observations, aquaplanet simulations, and theory. Observations are first used to identify the CRE on summertime meridional radiative gradients. Aquaplanet simulations then provide a controlled framework for isolating the dynamical importance of cloud-radiative effects and disentangling the physical pathways through which they influence storm activity. Finally, a minimal dynamical model is used to interpret the feedback in terms of coupled interactions among storms, clouds, and SST gradients and to demonstrate the emergent cloud properties that control its strength. Together, this hierarchy allows us to determine the role of CRE in storm-track seasonal maintenance and identify the physical processes that govern the strength of this feedback.

%% file: Results.tex
\section*{Results}
\subsection*{Observational evidence for cloud-radiative reinforcement of storm tracks}

\begin{figure}[t!]
    \centering
    \includegraphics[width=1\linewidth]{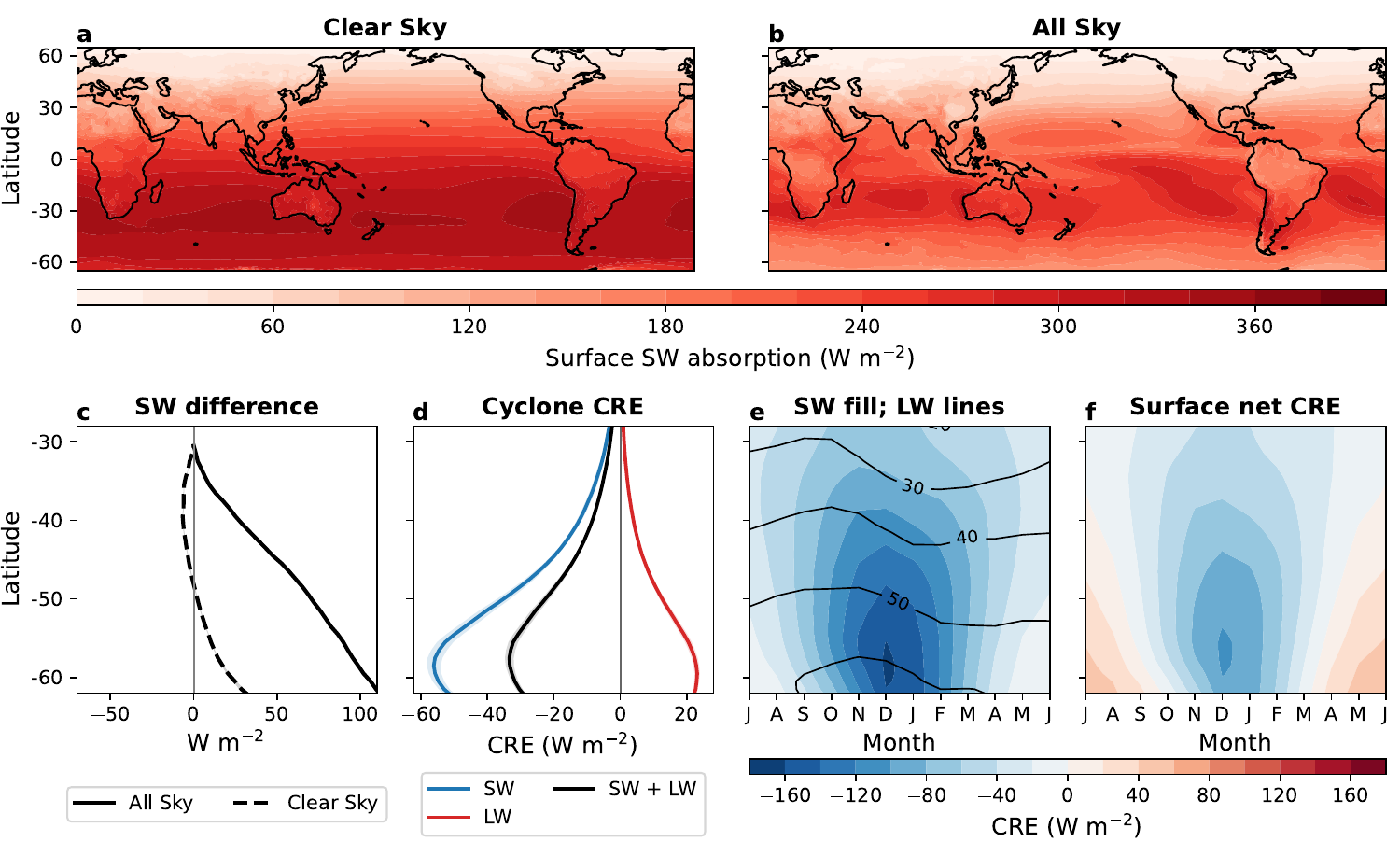}    
    \caption{\textbf{Observed summertime surface SW gradients and their modification by CRE}
(a,b) CERES surface SW absorption for clear-sky and all-sky conditions in DJF, respectively. 
(c) The zonal-mean change in surface SW absorption relative to latitude -30$^\circ$ for the SH during DJF under all-sky (solid) and clear-sky (dashed) conditions.
(d) SH DJF cyclone-associated surface CRE within 1,000 km of cyclone centers, showing shortwave (blue), longwave (red), and total (black) contributions, proportional to the CRE within cyclones times their frequency of occurrence.  
(e) Surface SW and LW (color and contours, respectively), and (f) net CRE as a function of month and latitude for the SH. Shading indicates 95\% CI (Methods). In panel c, CI is thinner than the line width.}
    \label{fig1}
\end{figure}

In this section, we use CERES observations (Methods) to identify a potential mechanism through which clouds may support summertime storm activity in the SH. Because summer insolation is large and is absorbed primarily at the surface \cite{stephens2012update}, we focus on the meridional gradient in absorbed surface SW radiation. Under clear-sky conditions, this gradient is weak across the SH midlatitudes (Fig.~\ref{fig1}a). The zonal-mean absorbed SW difference relative to 30$^\circ$ remains close to zero up to about 50$^\circ$ latitude and reaches only about 25 W m$^{-2}$ by 60$^\circ$ (Fig.~\ref{fig1}c). Such weak radiative forcing alone provides little support for meridional temperature gradients and summertime storm activity.

The inclusion of SW CRE substantially alters this radiative balance. Tracking individual cyclones and sampling CRE within 1,000~km of their centers reveals that storm-associated cloud radiative cooling is concentrated on the poleward flank of the SH storm track (Fig.\ref{fig1}d), strongly reducing surface SW absorption at high latitudes (Fig.\ref{fig1}b). This CRE generates meridional radiative gradients roughly four times larger than those under clear-sky conditions (Fig.~\ref{fig1}c). In effect, the cloud field creates a summertime radiative forcing that is largely absent in clear-sky conditions. Because these clouds are generated by the storms themselves, the observations point to a self-reinforcement mechanism: storms produce an extensive cloud field, the clouds preferentially cool high latitudes, and the resulting SST and temperature gradients reinforce subsequent storm activity. The seasonal and latitudinal structure of the CRE indicates that this meridional forcing would be only partly offset by the LW CRE, since it is more uniform across latitudes than SW CRE (Fig.~\ref{fig1}e contours versus color). Consequently, the net CRE retains the meridional structure imposed by SW CRE (Fig.~\ref{fig1}f). 

In the NH, the weaker summertime all sky contrast (Fig.\ref{fig:S1}) suggests less cloud-radiative reinforcement of storm activity. Extensive land cover modifies both sides of the radiative contrast: higher surface albedo reduces SW absorption near 30$^{\circ}$, while weaker cloud radiative cooling at 50–60$^{\circ}$ limits the poleward reduction in absorption. The resulting meridional contrast in absorbed surface SW radiation is about 40\% weaker than in the SH (Fig.\ref{fig:S1}c). This difference is consistent with stronger cloud-radiative support for the SH summer storm track. However, NH storm activity weakens already early in the warm season (Fig.~\ref{fig:S1}d), pointing to additional contributions from faster seasonal land warming, alongside the effects of topography and ocean circulation \cite{shaw2022stormier}.

Together, these observations identify a potential cloud-radiative feedback on the summertime SH storm track. The extensive cloud fields associated with early-summer storm activity strongly reduce high-latitude shortwave absorption, generating meridional radiative gradients that could reinforce the SST gradients that support subsequent storm activity. However, the dynamical importance of this mechanism cannot be inferred from observations alone, because summertime storm activity may also be influenced by other processes, including the maintenance of SST gradients by the ocean mixed layer's thermal inertia \cite{shaw2018moist} and seasonal changes in Southern Ocean circulation. To isolate the dynamical importance of CRE and disentangle the physical pathways through which it influences summertime storm activity, we next turn to idealized aquaplanet simulations.

\subsection*{Isolating the impact of midlatitude CRE using aquaplanet simulations}

\begin{figure}[t!]
    \centering
    \includegraphics[width=1\linewidth]{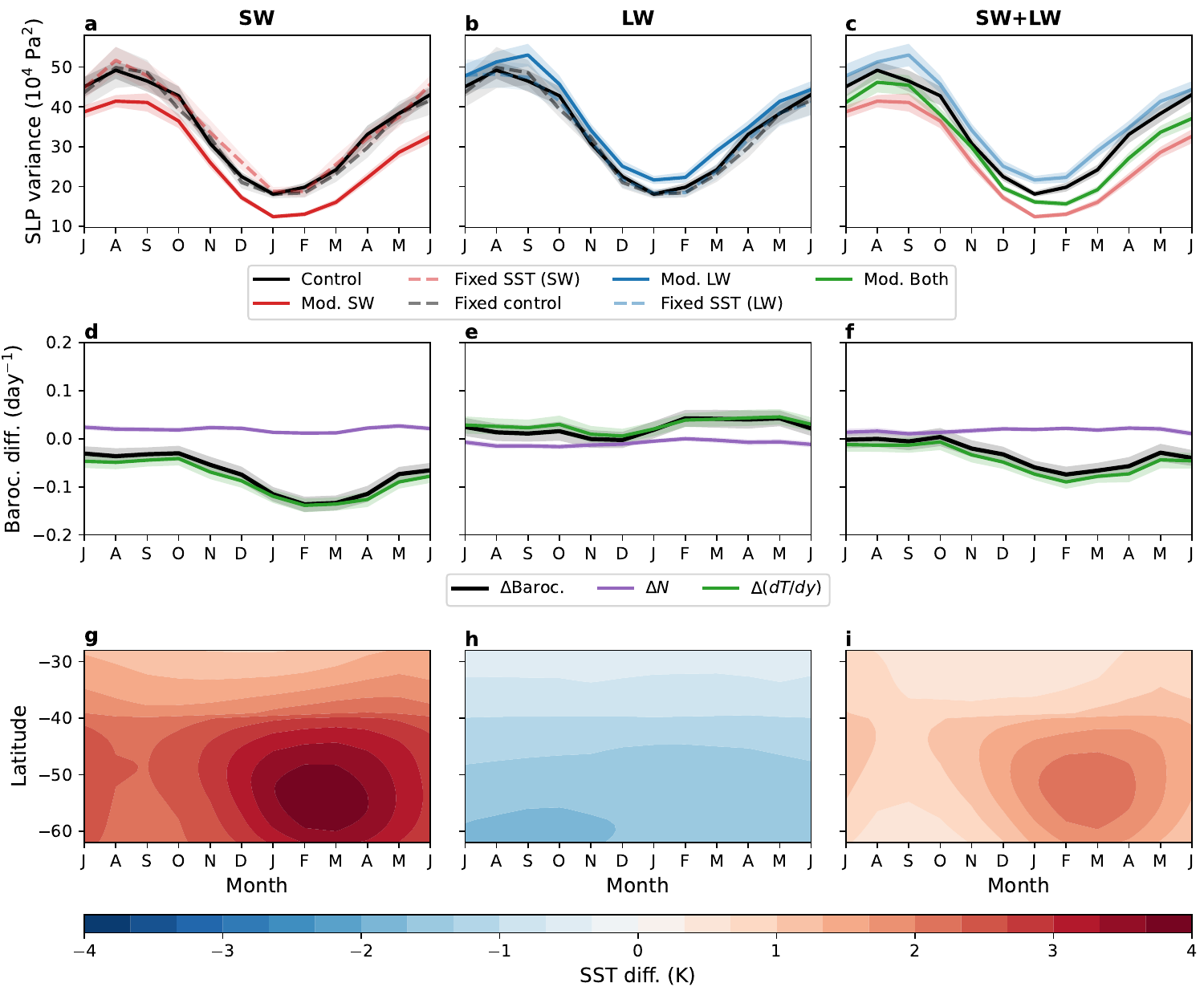}
    \caption{\textbf{Aquaplanet simulations reveal the role of CRE in maintaining summer storm tracks} (a) Monthly mean SLP variance, averaged zonally, between latitudes 30-60$^\circ$S, and multi-annually for the Control (black), modified SW (Mod. SW, red), control with fixed SST (gray dashed) and Mod. SW with fixed SST (red dashed). (d) The difference in the atmospheric baroclinicity (Baroc. diff., Eq.~\ref{eq:baroc}) between Mod. SW and Control as a function of month ($\Delta$Baroc., black), with the contribution from the vertical stratification ($\Delta$N, purple), and the meridional temperature gradient ($\Delta(dT/dy)$, green) (g) Zonal mean sea surface temperature difference (SST diff.) between the Mod. SW and control simulations as a function of month and latitude. Panels (b,e,h) and (c,f,i) are analogous to (a,d,g), but for the simulations with reduced LW CRE and with both reduced SW and LW CRE, respectively. Shading indicates 95\% CI (Methods). All SST changes are statistically significant ($p<0.05$).}
    \label{fig:aquaplanet_results}
\end{figure}

The observational analysis suggests that storm-associated surface SW CRE may help sustain storm activity by generating meridional radiative gradients that are largely absent under clear-sky conditions during summer. However, observations alone cannot establish whether this feedback is dynamically important, nor can they distinguish its contribution from those of LW CRE or direct atmospheric cloud-radiative heating. To isolate the impact of different cloud-radiative feedbacks and test whether they can indeed sustain storm activity, we perform a series of aquaplanet simulations in which SW and LW CRE are independently reduced by 25\% over the midlatitudes (Methods). We then examine the response of storm activity and its seasonality.

Reducing SW CRE substantially weakens the storm track (Fig.~\ref{fig:aquaplanet_results}a). The response ($\frac{\mathrm{Perturbed}-\mathrm{Control}}{\mathrm{Control}}$) is highly seasonal, with only a modest reduction in late winter and spring (August-October, ASO; 14$\pm$4\%) but a pronounced weakening in late summer and autumn (February–April, FMA; 33$\pm$4\%). This seasonality closely tracks changes in atmospheric baroclinicity, which is almost entirely explained by a weakening of the meridional temperature gradient, while changes in static stability act to slightly offset the response (Fig.~\ref{fig:aquaplanet_results}d). 
The weakened atmospheric temperature gradient can be directly traced to the SST response induced by cloud-radiation modification.
Reducing SW CRE increases surface SW absorption, producing a pronounced high-latitude warming during late summer and autumn (Fig.~\ref{fig:aquaplanet_results}g). The spatial structure of this warming closely follows the observed SW CRE distribution in observations (Fig.~\ref{fig1}e), while the seasonal lag arises from the large heat capacity of the ocean mixed layer. 
This warming reduces the equator-to-pole temperature contrast, thereby weakening baroclinic instability. 
In addition, a weaker year-round reduction in the SST gradient persists throughout the seasonal cycle, contributing to weaker storm activity year-round.
The weakening of storm activity is accompanied by reduced total cloud
cover over 50--60$^\circ$S, particularly during summer
(Fig.~\ref{fig:TCC_change}). This cloud response is consistent with
a positive feedback: the initial reduction in SW CRE weakens storm
activity, while the associated loss of cloud cover further reduces
high-latitude SW reflection, reinforcing the initial perturbation.

The response to LW CRE is qualitatively opposite but substantially weaker. Reducing LW CRE strengthens storm activity throughout the year (Fig.~\ref{fig:aquaplanet_results}b), consistent with an increase in baroclinicity driven by stronger meridional temperature gradients (Fig.~\ref{fig:aquaplanet_results}e), which originates from a modest high-latitude cooling (Fig.~\ref{fig:aquaplanet_results}h). The relatively uniform SST and storm activity responses are consistent with the weak seasonal cycle of observed surface LW CRE (Fig.~\ref{fig1}e contours versus color).

 When both SW and LW CRE are reduced simultaneously, the response closely follows the sum of the individual responses (monthly pattern correlations of 0.99 for SST and 0.84 for storm activity), with annual-mean additivity residuals ($\Delta_{\mathrm{SW+LW}} - \Delta_{\mathrm{SW}} - \Delta_{\mathrm{LW}}$) of $0.099 \pm 0.007$~K for SST and $-0.51 \pm 1.2$~hPa$^{2}$ for storm activity (Fig.~\ref{fig:aquaplanet_results}c,f,i). Thus, the strengthening induced by reduced LW CRE largely offsets the weaker year-round component of the SW CRE response, leaving a circulation change dominated by the strongly seasonal SW effect. As a result, the net response is concentrated in late summer and autumn (20$\pm5$\% reduction at FMA), while the annual-mean weakening is substantially reduced (21$\pm2$\% for the SW experiment versus 11$\pm2$\% for the all radiation experiment). The seasonal response is robust across storm metrics, perturbation magnitudes and cloud microphysics schemes, although its magnitude depends on the configuration (Methods; Figs.~\ref{fig:properties} and \ref{fig:sensitivity}).

To determine whether the storm-track response arises from direct atmospheric CRE or from SST changes induced by CRE, we perform an additional set of simulations in which SSTs are held fixed at the control climatology (see Methods, Fig.~\ref{fig:aquaplanet_results}a,b). In these experiments, the storm-track response is much smaller, indicating that midlatitude CRE-driven reinforcement of storm activity on seasonal timescales predominantly operates via SST rather than direct atmospheric heating. This result is consistent with previous studies showing that decoupling clouds from the midlatitude circulation, which mainly alter atmospheric CRE due to the "smoothing" effect of the surface, produces only modest changes in storm activity \cite{Grise2019}, and that confining atmospheric CRE perturbations to the midlatitude significantly reduces the circulation response \cite{albern2020role}. The apparent discrepancy with previous atmospheric CRE studies (e.g., \cite{li2015influence}) arises from our weaker (25\% rather than 100\%) and midlatitude-confined CRE reduction, which produces an average (30-60$^\circ$, 850-300~hPa) thermal anomaly magnitude 6.6 times smaller (Fig.~\ref{fig:ACRE}). 

To assess whether the proposed mechanism is consistent with the behavior of comprehensive climate models, we examine the CMIP6 \cite{eyring2016overview} intermodel spread along the physical pathway \cite{chen2020emergent,schiro2022model} linking Storm-track CRE (50-60S$^\circ$ average), meridional surface radiative forcing, SST gradients, atmospheric baroclinicity, and storm activity (Fig.~\ref{fig:CMIP6}). We relate DJF (JJA) radiative forcing to FMA (ASO) SST and circulation responses to account for the delayed ocean adjustment evident in the aquaplanet simulations. During late summer and autumn (February–April, FMA), all four successive relationships exhibit statistically significant correlations ($p<0.01$). The pairwise intermodel relationships are:
\[
\mathrm{CRE}
\underbrace{\longleftrightarrow}_{R^2=0.62}
\Delta R
\underbrace{\longleftrightarrow}_{R^2=0.55}
\Delta\mathrm{SST}
\underbrace{\longleftrightarrow}_{R^2=0.47}
\mathrm{Baroclinicity}
\underbrace{\longleftrightarrow}_{R^2=0.35}
\mathrm{Storm\ activity},
\]
where $\Delta R$ is the meridional surface radiative contrast. In late winter and spring (August–October), CRE explains only 14\% of the intermodel variance in radiative contrast, while subsequent relationships remain similar, consistent with the weaker cloud-radiative influence predicted by the aquaplanet experiments. However, substantial unexplained variability at each stage, which arises from processes such as subtropical cloud effects on the radiative gradient \cite{vcrnivec2023evaluating}, ocean dynamics effects on SST \cite{barpanda2020surface,shaw2022stormier}, and tropical forcing effects on baroclinicity \cite{kang2025robust}, obscures the relationship between midlatitude CRE and storm activity.  The direct correlation between CRE and storm activity, -0.3 ($p=0.13$) for summer and 0.4 ($p=0.03$) for winter, is thus more likely to result from the effect of increased storm activity on cloudiness \cite{Hadas2023}. This highlights the limitations of using the CMIP6 intermodel spread for isolating the feedback and motivates the use of controlled simulations to establish the proposed mechanism.

Taken together, these results demonstrate that midlatitude clouds reinforce storm activity on seasonal timescales primarily through their SW influence on SST gradients. Reducing SW CRE weakens high-latitude summer cooling, diminishing meridional SST gradients and baroclinicity, with the strongest storm-track weakening in late summer and autumn. Reducing LW CRE partly offsets the weaker year-round response but leaves this seasonal maximum intact. The much smaller response with fixed SSTs suggests that the reinforcement operates mainly through the surface rather than through direct atmospheric cloud-radiative heating.

\subsection*{A minimal theory for the cloud-radiative feedback}

The observational and aquaplanet analyses suggest that the storm track can self-reinforce through a feedback involving CRE and SST gradients. To understand why this feedback is effective and which cloud properties govern its strength, we develop a minimal dynamical framework that couples storm activity, meridional SST gradients, and CRE, intended to clarify the mechanism rather than to provide quantitative predictions. We focus on the evolution of the meridional SST difference in the midlatitudes ($\Delta T = T_{30}-T_{60}$). Neglecting ocean heat transport, consistent with the slab-ocean configuration used in the aquaplanet experiments, we approximate the evolution of $\Delta T$ as a balance between differential solar heating, atmospheric storms' heat transport, and differential outgoing longwave radiation (OLR) \cite{barsugli1998basic}:
\begin{align}
    C_o \frac{d\Delta T}{dt}
    =
    -\underbrace{C_a\frac{F_e}{\Delta y}}_{\text{storms heat transport}}
    +
        \underbrace{\Delta\mathrm{SW}}_{\text{SW radiative forcing}}
-\underbrace{\Delta\mathrm{OLR}}_{\text{OLR damping}},
    \label{eq:model_dt_general}
\end{align}
where $C_o$\ and $C_a$\, are the effective oceanic and atmospheric heat capacities per unit area, $F_e$ is the meridional heat transport by storms (units of  K m s\(^{-1}\)), $\Delta y=3,000$~km is the approximate meridional distance between latitude 30$^\circ$ and 60$^\circ$.  The heat capacities are estimated using column estimates:
\begin{align}
    C_a \simeq \frac{c_p p_s}{g} \simeq 10^7~\mathrm{J\,m^{-2}\,K^{-1}},
    \qquad
    C_o \simeq \rho_w c_w h \simeq10^8~\mathrm{J\,m^{-2}\,K^{-1}},\label{eq:Co}
\end{align}
where $c_p$ is the specific heat capacity of air,
$p_s$ is surface pressure, $g$ is gravitational acceleration,
$\rho_w$ and $c_w$ are the density and specific heat capacity of seawater,
respectively, and $h=30$\,m is the aquaplanet mixed-layer depth.

To close the system, storm heat transport is represented using a down-gradient closure \cite{Green1970,Vallis2017atmospheric}. Assuming that storms adjust to the large-scale temperature gradient on a synoptic timescale of $\tau_s=1$ week, the heat transport is expressed as:
\begin{align}
    \tau _s\frac{dF_e}{dt} = \kappa \frac{\Delta T}{\Delta y} -F_e,
    \label{eq:eddies}
\end{align}
where $\kappa=\frac{\Delta y^2}{\tau_s}=1.5\times10^7~$m$^2$~s$^{-1}$ is the eddy diffusivity, set by a dimensional closure motivated by the mixing-length form of eddy-transport parameterizations. This formulation represents the tendency of storms to relax meridional temperature gradients through poleward heat transport, while adjusting rapidly relative to the seasonal evolution of the SST field. $F_e $ (K~m~s\(^{-1}\)) is used as a measure of storm activity because it directly enters the thermal budget and, like SLP variance, scales with the magnitude of synoptic eddies.

The SW radiative forcing is separated into clear-sky and cloud-radiative contributions. The small ocean surface albedo is neglected, treating the surface as perfectly absorbing and eliminating multiple reflections between the surface and cloud base. In the absence of cloud effects, the absorbed SW contrast is
\begin{align}
    \Delta\mathrm{SW}_{\rm clear}(t)
    = I_{30}(t)-I_{60}(t),
\end{align}
where $I_{30}$ and $I_{60}$ are the incoming clear-sky SW fluxes at the surface at 30$^\circ$ and 60$^\circ$ latitude. This contrast is strongly seasonal, providing a large radiative tendency to maintain $\Delta T$ in winter but virtually vanishing during summer (Fig.~\ref{fig:1}b red). The clear-sky model therefore reproduces the central puzzle identified in the introduction: in the absence of CRE, there is little radiative forcing available to sustain summertime storm activity.

Because SW CRE dominates the seasonal storm-track response, we retain its dependence on storm activity while neglecting LW CRE. Storm-associated clouds preferentially reflect sunlight on the poleward side of the storm track. Motivated by observations showing that cloudiness saturates at high storm activity \cite{Hadas2023}, we set $\alpha_{30}=0$ and $\alpha_{60}(F_e)=\alpha_c\sigma(cF_e)$, giving
\begin{align}
    \Delta\mathrm{SW}(t,F_e)
    = I_{30}(t)
    -\left[1-\alpha_c\sigma(cF_e)\right]I_{60}(t),
    \label{eq:sw_contrast}
\end{align}
where $\alpha_c$ represents the maximum cloud albedo attainable over the storm track, $\sigma(x)=2[1+\exp(-x)]^{-1}-1=\tanh(x/2)$
is a sigmoid function satisfying $\sigma(0)=0$ and
$\sigma(x)\to1$ as $x\to\infty$, and $c$ sets the response of the cloud field to storm activity. Previous analysis of ERA-5 reanalysis, CERES observations, and midlatitude storm tracks estimated $\alpha_c$ to be approximately 0.3 in the current climate \cite{Hadas2023}. $c$ was set by default to $0.05~\mathrm{s\,m^{-1}\,K^{-1}}$. Stronger storms therefore increase the poleward albedo.

Differential heating that is not balanced by atmospheric heat transport
can instead be offset by differential energy loss to space through OLR.
As the atmosphere adjusts to the underlying SSTs, the warmer equatorward
column emits more longwave radiation to space than the colder poleward
column, damping the meridional temperature contrast. We represent this Planck response by linearizing the OLR contrast with respect to the SST contrast:
\begin{align}
    \Delta\mathrm{OLR}\simeq\lambda_{\mathrm{OLR}}\Delta T,
    \qquad
    \label{eq:OLR_damping}
\end{align}
where $\lambda_{\mathrm{OLR}}=3~\mathrm{W\,m^{-2}\,K^{-1}}$ is chosen to represent the magnitude of Planck damping \cite{soden2006assessment}. This effective closure captures OLR's tendency to oppose differential heating \cite{shaw2018moist}.

\begin{figure}
    \centering
    \includegraphics[width=1\linewidth]{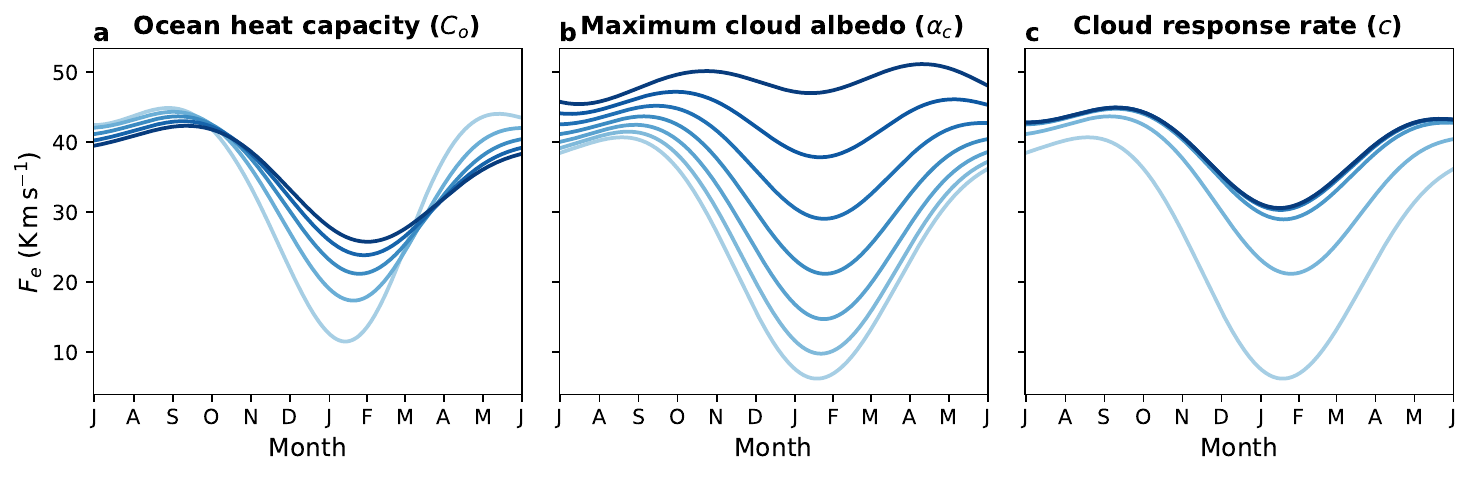}
    \caption{\textbf{Conceptual model for the cloud-radiative reinforcement of summer storm tracks} (a) Storms' heat flux ($F_e$,  Kms\(^{-1}\)) obtained from Eqs.~\ref{eq:model_dt_general}\&\ref{eq:eddies} for ocean heat capacities $C_o=0.5$--$1.5\times10^8$~J\,m$^{-2}$\,K$^{-1}$ in increments of $0.25\times10^8$ (light to dark blue).
(b) Same as (a), but varying maximum cloud albedo $\alpha_c$ from 0 to 0.6 in increments of 0.1 (light to dark blue).
(c) Same as (a), but varying cloud sensitivity $c$ from 0 to 0.25 in increments of 0.05 ($\mathrm{s\,m^{-1}\,K^{-1}}$, light to dark blue).}
    \label{fig:model}
\end{figure}

Next, the model's sensitivity to the two parameters that characterize the cloud feedback is examined. The sensitivity to $\alpha_c$ provides an idealized representation of the SW CRE perturbation applied in the aquaplanet experiments.  We vary $\alpha_c$ from 0, representing the absence of cloud albedo, to 0.6, about twice its present-day value. Increasing $\alpha_c$ strengthens storm activity throughout the year while substantially reducing its seasonal cycle (Fig.~\ref{fig:model}b). In the absence of cloud reflection ($\alpha_c=0$), summer storm activity is less than a quarter of its winter value. As $\alpha_c$ increases, enhanced high-latitude reflection strengthens the summertime SST gradient, progressively increasing summer storm activity. For sufficiently large $\alpha_c$, the cloud-induced radiative forcing largely compensates for the seasonal reduction in clear-sky meridional gradient, and the seasonal cycle in storm activity nearly disappears.

The sensitivity of cloud albedo to storm activity highlights a second key aspect of the feedback (Fig.~\ref{fig:model}c). Because cloudiness responds nonlinearly to storm activity and cloud reflection feeds back positively onto storm growth, the system exhibits a rapid transition between two distinct regimes. For weak sensitivities, CRE remains small, and storm activity retains a pronounced seasonal cycle. Once the cloud response exceeds a threshold, the cloud field rapidly approaches saturation during summer, leading to intense storm activity. In this conceptual model, the feedback therefore acts as a switch between weak- and strong-summer-storm regimes: relatively modest changes in the cloud response to storm activity can produce rapid changes in storm activity seasonality.

This framework provides a simple way to interpret how cloud representations influence storm-track behavior. Processes that alter either the maximum attainable cloud albedo or the sensitivity of cloudiness to storms, including cloud microphysics, convection, model resolution, and moisture availability, directly modify the strength of the feedback \cite{blanco2025insights}. The framework therefore provides a physically motivated basis for understanding how cloud biases can propagate into biases in storm-track strength and seasonality.

%% file: Conclusions.tex
\section*{Discussion}

Traditionally, storm tracks were viewed as being maintained by externally imposed meridional temperature gradients \cite{charney1947,Eady1949}. This picture was later revised to recognize that moist processes, particularly latent heat release, can also help maintain the thermal structure that supports storm growth \cite{Hoskins1990,Papritz2015,auestad2026latent}. Our results suggest that CREs constitute an additional maintenance mechanism on seasonal timescales. We show that early-summer storm-generated cloud fields create strong SW radiative gradients that reinforce meridional temperature gradients, thereby enhancing storm activity during late summer and autumn. Unlike latent heating, however, this mechanism is not confined to the atmosphere. Instead, it operates through coupled interactions among storms, clouds, and the ocean, with cloud-radiative forcing modifying SST gradients that subsequently feed back onto storm activity.

A key implication of the ocean-mediated nature of this mechanism is that its influence on storm activity is difficult to infer from the evolution of individual storms. A back-of-the-envelope estimate illustrates why. If storm clouds produce an albedo contrast of order 0.3 \cite{Hadas2023} under an incoming SW flux of 400 W m$^{-2}$, the resulting differential SW forcing is about 120 W m$^{-2}$. For a storm-associated cloud-radiative anomaly with a characteristic width of approximately 1000 km propagating at 10 m s$^{-1}$, the local forcing time is about one day. The resulting energy input is therefore $Q \sim 120 \times 86400 \approx 1\times10^{7}~\mathrm{J~m^{-2}}$. For a 30 m ocean mixed layer, the heat capacity is $C_o \approx 10^{8}~\mathrm{J~m^{-2}~K^{-1}}$ (Eq.~\ref{eq:Co}), corresponding to an SST change of only $\Delta T = Q/C_o \approx 0.09~\mathrm{K}$. Such a small SST perturbation is unlikely to substantially alter the life cycle of the storm that produced it. This does not preclude direct atmospheric CRE effects on individual storms, which can modify their thermal structure and baroclinic growth \cite{keshtgar2022cloud}. However, atmospheric CRE effects depend on cloud configuration and may cancel across storms \cite{voigt2023tug}, potentially leading to cancellation when averaged over many storms. Consequently, understanding the surface-mediated cloud-circulation feedback requires integrating CREs across the lifetimes of many storms and relating them to the evolving SST field.

Another important implication of this work concerns the interpretation of high-resolution climate simulations. Increasing model resolution often produces substantial changes in cloud structure, cloud fraction, and cloud-radiative effects \cite{stevens2019dyamond}, making such simulations a powerful tool for understanding cloud-circulation interactions. However, the computational cost of these experiments frequently necessitates short integrations and the use of prescribed SST configurations. Because the dominant cloud feedback identified here is mediated by SSTs and emerges over timescales of several months, high-resolution simulations with prescribed SSTs may accurately diagnose how resolution changes clouds but substantially underestimate how those cloud changes influence storm tracks. Capturing this interaction, therefore, requires simulations that integrate interactive SSTs over timescales long enough for the ocean surface to adjust.

A broader implication of this framework is that understanding cloud-circulation coupling may require moving beyond traditional diagnostics based on the mean properties of cloud fields. The mechanism identified here depends not only on the magnitude of the CRE but also on how storm activity organizes its meridional and seasonal structure. This suggests that developing a mechanistic understanding of the relationship between storm activity and the resulting distribution of CRE may provide a more physically meaningful diagnostic of cloud-circulation coupling and help explain how biases in cloud representation translate into biases in storm-track strength and seasonality.

The results presented here point to ocean dynamics as the next major challenge in understanding the feedback of midlatitude clouds on the circulation. While prescribing ocean properties such as mixed-layer depth and ocean heat fluxes reveals crucial insights into the magnitude and seasonality of storm tracks here and in previous studies \cite{barpanda2020surface,shaw2022stormier}, it neglects the dynamical response of the ocean to the storm activity: ocean circulation and mixed-layer depth are directly affected by storm activity \cite{oke2004oceanic,kang2020walker,du2026southern}, and changes in these properties can alter the evolution of SST gradients, thus altering the cloud-radiative feedback identified here. The ocean is therefore not simply a medium through which the feedback operates, but an active participant in the coupled storm-cloud-ocean system. Understanding how ocean dynamics regulate and are affected by this interaction represents a crucial next step.

%% file: Methods.tex
\section*{Methods}
\subsubsection*{ERA-5 reanalysis and CERES observations}

Observational estimates of storm activity are obtained based on 3-hourly sea level pressure (SLP) from the European Centre for Medium-Range Weather Forecasts (ECMWF) ERA5 reanalysis \cite{Hersbach2020}. 

Radiative fluxes and cloud radiative effects are obtained from the Clouds and the Earth's Radiant Energy System (CERES) SYN1deg product \cite{wielicki1996clouds,rutan2015ceres}. The dataset provides surface and top-of-atmosphere radiative fluxes on a $1^\circ \times 1^\circ$ grid at hourly resolution, allowing diagnosis of the seasonal and spatial structure of cloud radiative forcing.

Storm-associated CRE (Fig.~\ref{fig1}d) is diagnosed using SLP based cyclone trajectories identified with the TRACK algorithm (\cite{hodges_1995}, see \cite{hadas2025lagrangian} for the specific configurations). At each three-hourly track position, we sample surface CRE within 1,000~km of the cyclone centre, including all stages of the cyclone lifetime. Cyclone CRE footprints are averaged over longitude and time to obtain DJF latitude profiles. The resulting diagnostic therefore includes both cyclone occurrence and the CRE within cyclone footprints. 

\subsection*{Midlatitude storm activity}

Storm activity is quantified as the variance of 2.5--6-day
bandpass-filtered SLP anomalies, normalized by
$\sin^2\phi$, where $\phi$ is latitude. This normalization accounts
for the latitude dependence of the Coriolis parameter: under geostrophic
balance, a given pressure gradient corresponds to weaker winds at
higher latitudes. Storm activity is averaged zonally and area weighted between latitudes 30-60$^\circ$.

\subsection*{Aquaplanet simulations} \label{Simulations}

To isolate the impact of CRE, we perform idealized aquaplanet simulations with the Community Earth System Model version 2 (CESM2, \cite{danabasoglu2020community}), using the Community Atmosphere Model version 5 (CAM5, \cite{neale2010description}). The simulations are based on the QSC5 aquaplanet configuration with a 30 m slab ocean, but are modified to include a seasonal insolation cycle. QSC5 is used instead of QSC6 (the CAM6 version) because QSC6 produces excessive subtropical and midlatitude low-cloud cover when run with a seasonal cycle. The simulations use the f09 horizontal resolution, corresponding to approximately $0.9^\circ \times 1.25^\circ$ in latitude and longitude. Each simulation is integrated to statistical equilibrium and then continued for an additional 30 years, which are used for analysis. To make the simulation more representative of Southern Hemisphere conditions without explicitly simulating sea ice, thereby reducing model complexity, surface evaporation is suppressed, and the surface albedo is fixed at 0.7 poleward of $65^\circ$ latitude, mimicking a cold, highly reflective Antarctic-like surface. Although the aquaplanet simulation does not perfectly reproduce the observed CREs, it captures the key features relevant to this study: strong summertime SW CRE on the poleward flank of the storm track and weaker, more spatially and temporally uniform LW CRE (Fig.~\ref{fig:S2}).

A central element of the experimental design is a controlled modification of the CRE over the midlatitude. Rather than altering cloud properties themselves, we directly reduce their radiative influence by blending all-sky radiative quantities toward their corresponding clear-sky values:
\begin{equation}
Q' = Q_{\mathrm{clr}} + s \left( Q_{\mathrm{all}} - Q_{\mathrm{clr}} \right),
\end{equation}
where $Q_{\mathrm{all}}$ and $Q_{\mathrm{clr}}$ denote the all-sky and clear-sky radiative quantities, respectively. The scaling is applied consistently to both atmospheric radiative heating rates and surface radiative fluxes, thereby modifying the temperature tendencies of both the atmosphere and the slab ocean. The parameter $s$ controls the strength of the cloud radiative effect. A value of $s=1$ corresponds to the standard model, while smaller values weaken the impact of clouds on radiation, approaching a clear-sky limit as $s \to 0$. The modification is applied only poleward of $40^\circ$ latitude in both hemispheres, where most storm-associated CRE is concentrated (Fig.~\ref{fig1}d).  

Removing the net cloud radiative effect in simulations with interactive SSTs can produce substantial warming \cite{Voigt2021}, motivating the use of prescribed SSTs in previous experiments that remove CRE, such as COOKIE \cite{stevens2012cookie}. Our experimental design addresses this difficulty in two ways. First, we reduce CRE by only 25\%, thereby limiting the thermal response. This partial reduction also allows changes in cloudiness to feed back into the circulation, a coupling that is excluded by cloud locking and COOKIE experiments. Second, in the interactive-SST experiments, we apply a spatially uniform ocean Q-flux, determined separately for each perturbation, to maintain the climatological equatorial SST at its control value. This adjustment minimizes tropical SST changes and associated remote circulation responses that could influence midlatitude storm tracks \cite{Kang2008}, without directly imposing meridional gradients in ocean heating.

We conduct a control simulation with unmodified CRE (s=1) and three sensitivity experiments that reduce SW CRE alone, LW CRE alone, or both by 25\% (s=0.75). These experiments isolate the individual and combined effects of SW and LW CRE on storm activity. To separate the direct atmospheric response from changes mediated by surface temperature, we repeat the CRE-modification experiments with SST fixed to the climatology of the control simulation. The magnitude of the ocean heat flux was determined iteratively for each experiment, yielding cooling tendencies of 5.99 and 2.94~W~m$^{-2}$ for the SW and combined SW+LW perturbations, respectively, and a heating tendency of 3.17~W~m$^{-2}$ for the LW perturbation. The global mean heating (cooling) is found to weaken (strengthen) storm activity (e.g., Fig.~\ref{fig:sensitivity}c), consistent with previous studies \cite{Ogorman2008energy}.

The seasonal response reported in the main text using SLP variance is also evident in eddy kinetic energy, an alternative measure of storm activity based on synoptic-scale wind anomalies (Fig.~\ref{fig:properties}a). From the storm-population perspective, the weakening primarily reflects reduced intensity (Fig.~\ref{fig:properties}c), with relatively little change in storm numbers (Fig.~\ref{fig:properties}b). Longer storm lifetimes partly offset this weakening (Fig.~\ref{fig:properties}d), consistent with previous findings that storm growth and lifetime are negatively correlated with baroclinicity \cite{Schemm2018,hadas2025lagrangian}. Reducing net CRE by 12.5\% produces approximately half the FMA response (11$\pm5$\%, Fig.~\ref{fig:sensitivity}a). Repeating the 25\% CRE reduction experiment with the Rasch–Kristjánsson microphysics scheme instead of the default Morrison–Gettelman scheme also yields a similar seasonal response, with an FMA response of 15$\pm5$\% (Fig.~\ref{fig:sensitivity}b), showing the mechanism is robust under the usage of different microphysics schemes, although the magnitude of the effect might differ. These tests support the robustness of the mechanism and its seasonal structure, although the quantitative response remains specific to the model configurations examined here.

To connect changes in storm activity to the large-scale thermal structure of the atmosphere, we quantify the atmospheric baroclinicity using the Eady Growth Rate, a widely used measure of the atmosphere's capacity to support baroclinic growth \cite{Eady1949,Lorenz1955,Hadas2026stronger}:
\begin{equation}
\begin{aligned}
   \text{Baroc.} &= \frac{g}{NT}\frac{\Delta T}{\Delta y},\\
   N &= \sqrt{-\frac{\rho g^2}{\theta}\frac{\Delta\theta}{\Delta p}}.
\end{aligned}
\label{eq:baroc}
\end{equation}
where $g$ is the gravitational acceleration, $T$ is the temperature, $\rho$ is the air density, $\theta$ is the potential temperature, and $N$ is the Brunt–Väisälä frequency, which is a measure of the vertical stratification of the atmosphere. Baroc. increases with stronger meridional temperature gradients and weaker atmospheric stratification. The meridional temperature gradient is defined as the temperature difference at 850 hPa between latitudes -30$^\circ$ and -60$^\circ$. The vertical gradient in $\theta$ is taken as the difference between the $\theta$ at 300~hPa and 850~hPa at latitude -45$^\circ$.

\subsection*{Statistical testing}

The 95\% confidence intervals, shown as shading around curves or
reported as $\pm$ values, are calculated as
\[
\bar{x} \pm t_{0.975,n-1}\frac{s}{\sqrt{n}},
\]
where $\bar{x}$ is the climatological mean, $s$ is the standard
deviation based on interannual variability, $n$ is the number of
years, and $t_{0.975,n-1}$ is the 97.5th percentile of Student's
$t$ distribution with $n-1$ degrees of freedom.

%% file: Acknowledgment.tex
\subsection*{Acknowledgments}
\textbf{Funding:} This work was supported by the Azrieli Fellowship. \textbf{Author contributions:} O.H. conceived the study, developed the methodology, performed the simulations and analyses, interpreted the results, prepared the figures, and wrote the manuscript. \textbf{Competing interests:} The author declares that they have no competing interests. \textbf{Data and materials availability:} All data needed to evaluate the conclusions in the paper are present in the paper and/or the Supplementary Materials. ERA5 sea level pressure data are available through the Copernicus Climate Data Store \cite{Hersbach2023SingleLevels}. CERES SYN1deg radiative fluxes are available from the NASA Langley Atmospheric Science Data Center \cite{CERES_SYN1deg}. The CESM2 source code is publicly available through the ESCOMP GitHub repository \cite{CESM_GitHub}. The radiation scheme files used to implement the cloud-radiative effect modifications in the CESM2 simulations are available on Zenodo \cite{hadas2026radiation}.

%% file: Supplementary.tex
\setcounter{figure}{0}
\renewcommand{\thefigure}{S\arabic{figure}}

\section*{Supplementary figures}

\begin{figure}[b!]
    \centering
    \includegraphics[width=1\linewidth]{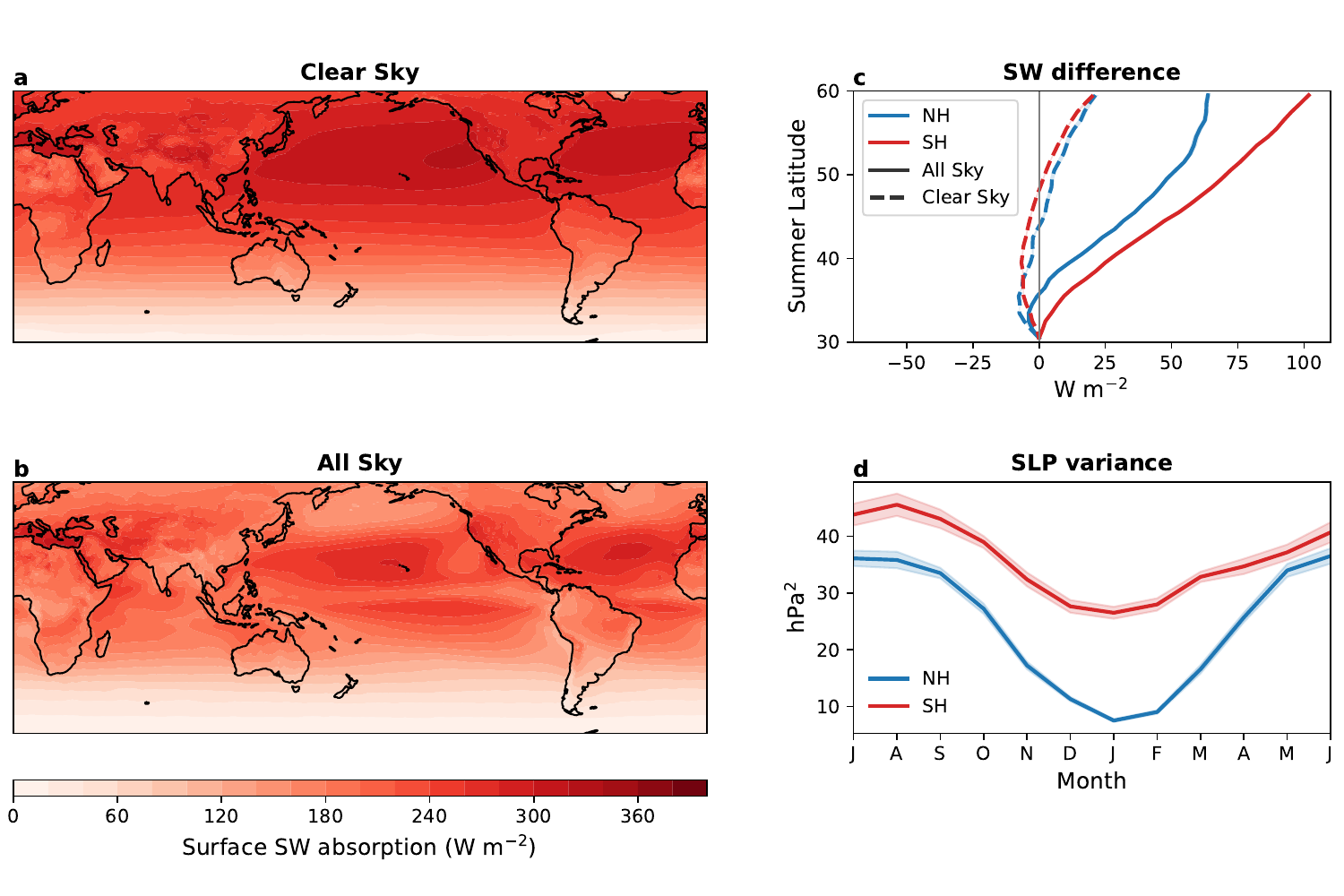}    
    \caption{\textbf{Cloud effect in summer is suppressed in the Northern Hemisphere}
(a,b) CERES surface shortwave absorption for clear-sky and all-sky conditions in JJA, respectively. (c) The zonal-mean change in surface shortwave absorption relative to latitude 30$^\circ$ for the NH (blue) and SH (red) under all-sky (solid) and clear-sky (dashed) conditions. (d) SLP variance for the NH and SH as a function of month. NH months were shifted by six months to match the SH. Shading indicates 95\% CI (Methods).}
    \label{fig:S1}
\end{figure}

\begin{figure}
    \centering
    \includegraphics[width=1\linewidth]{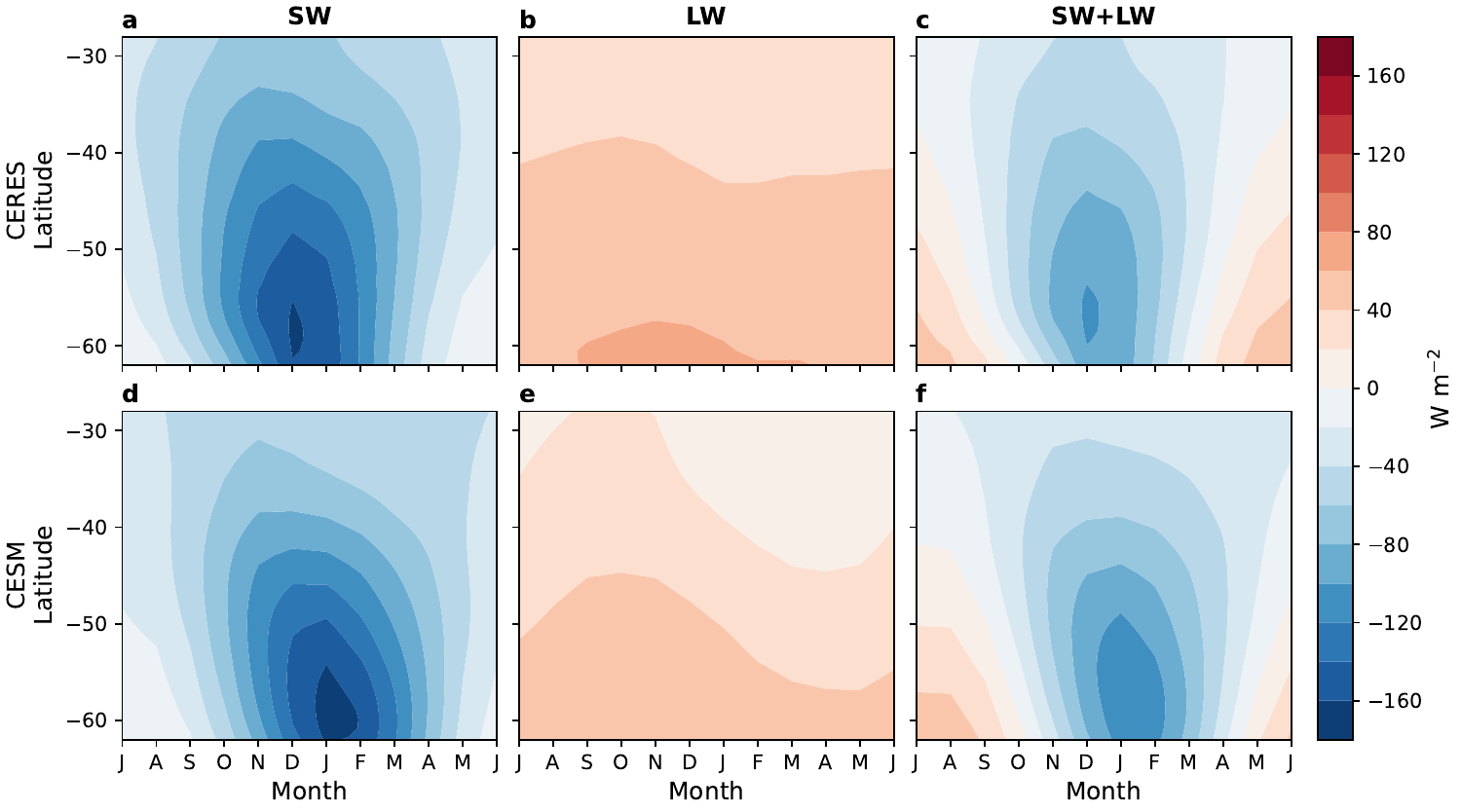}    
    \caption{\textbf{Comparison of observed and simulated cloud radiative effects}
(a-c) Surface shortwave, longwave, and net cloud radiative effects from CERES as a function of month and latitude in the Southern Hemisphere. (d-f) Corresponding fields from the CESM2 aquaplanet control simulation.}
    \label{fig:S2}
\end{figure}

\begin{figure}
    \centering
    \includegraphics[width=0.5\linewidth]{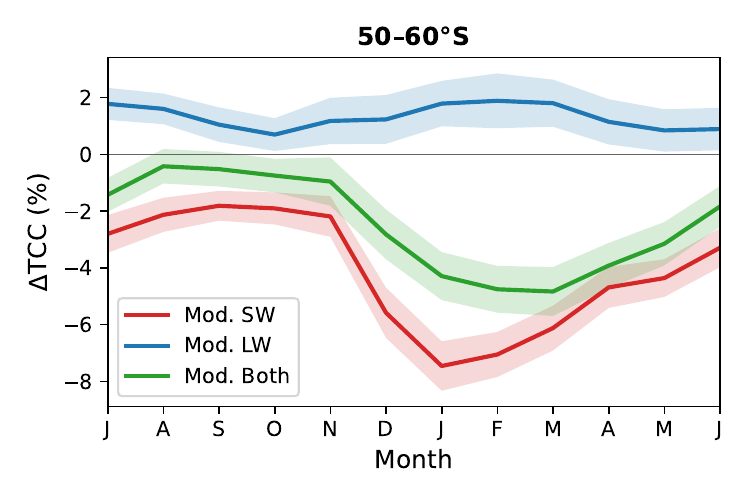}
    \caption{\textbf{Changes in storm track cloudiness are consistent with changes in storm activity} Zonal-mean and 50-60$^\circ$S mean total cloud-cover response to modified shortwave (Mod. SW), longwave (Mod. LW), and combined (Mod. Both) cloud-radiative effects relative to the control simulation as a function of month. Shading indicates 95\% CI (Methods).}
    \label{fig:TCC_change}
\end{figure}

\begin{figure}
    \centering
    \includegraphics[width=1\linewidth]{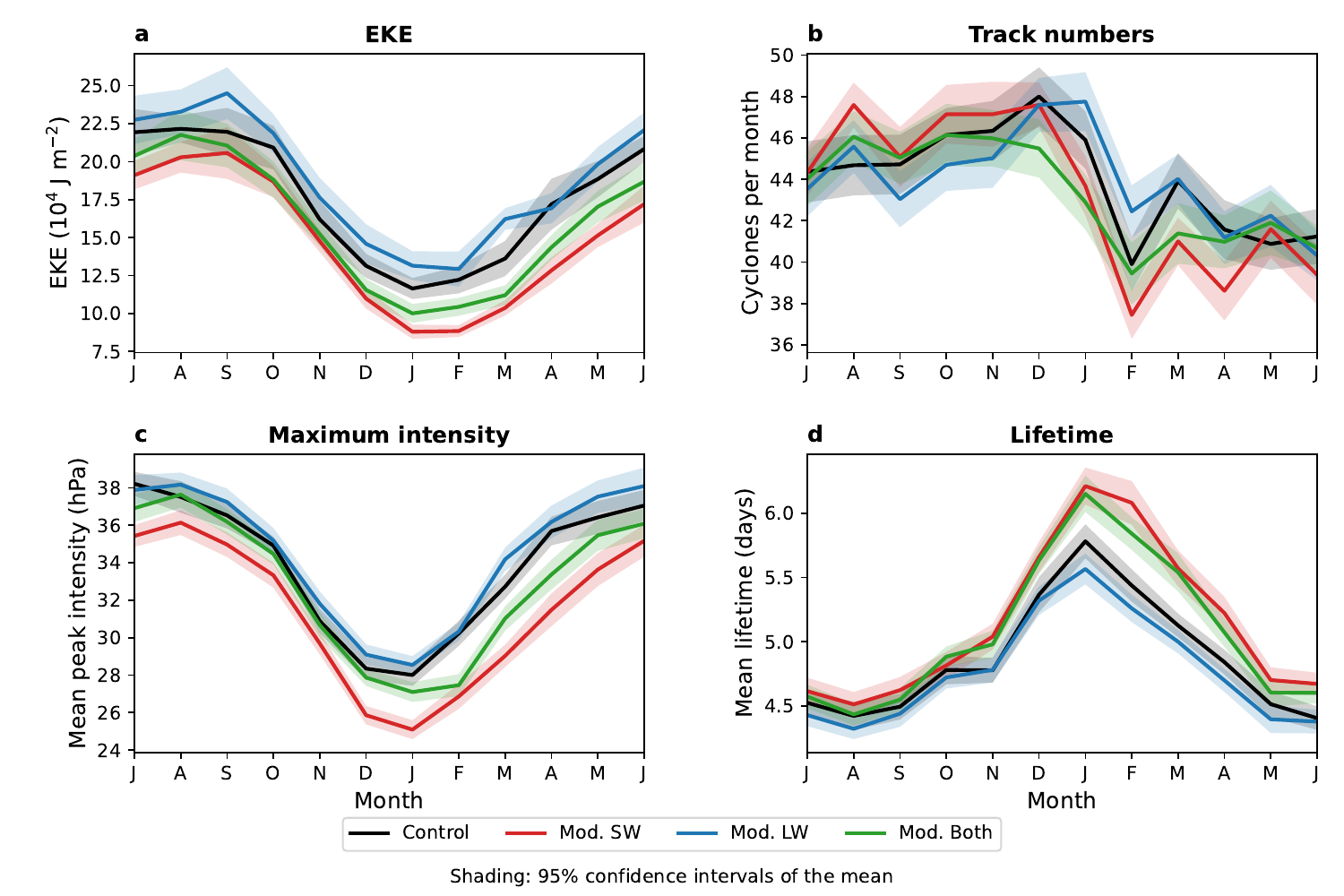}    
    \caption{\textbf{Different metrics for storm activity} (a) vertically integrated eddy kinetic
energy, $\mathrm{EKE}=-(2g)^{-1}\int_{850}^{250}
(u'^2 + v'^2)\,\mathrm{d}p$, averaged over $30$--$60^\circ$S, where $u'$
and $v'$ are 2.5-6 days band-pass wind perturbations, and $g$ is
gravitational acceleration (b) cyclone
count; (c) mean peak intensity; and (d) mean lifetime. Shading indicates 95\% CI (Methods).}
    \label{fig:properties}
\end{figure}

\begin{figure}
    \centering
    \includegraphics[width=\linewidth]{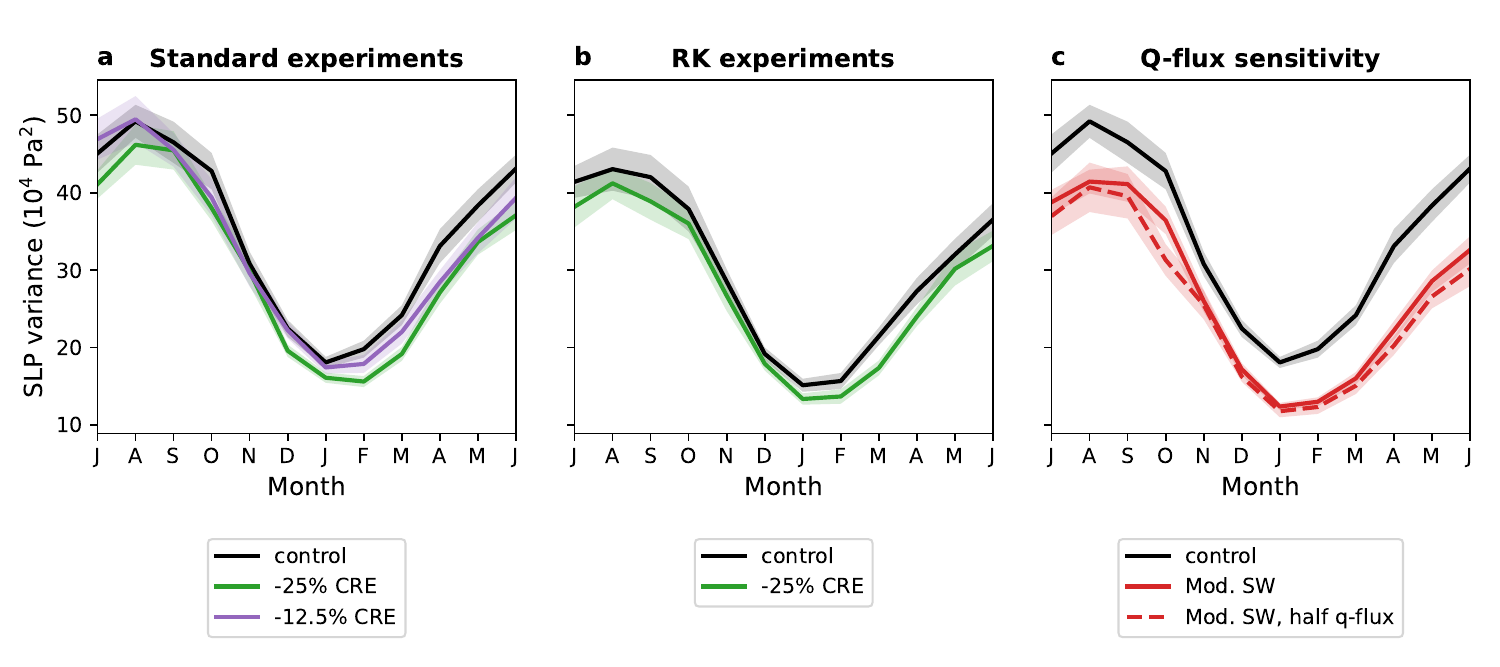}    
    \caption{\textbf{Sensitivity of the seasonal cycle of storm activity to CRE reduction and cloud microphysics parameterization} Monthly mean SLP variance averaged over 30$^\circ$–60$^\circ$S for (a) the control simulation and simulations with net CRE reduced by 25\% and 12.5\%, and (b) the control and 25\% CRE reduction experiments using the Rasch–Kristjánsson (RK) cloud microphysics scheme instead of the default Morrison–Gettelman (MG2) scheme. (c) Comparing control, Mod. SW, and Mod. SW with half the Q-flux. Shading indicates 95\% confidence intervals.}
    \label{fig:sensitivity}
\end{figure}

\begin{figure}[h]
    \centering
    \includegraphics[width=0.85\linewidth]{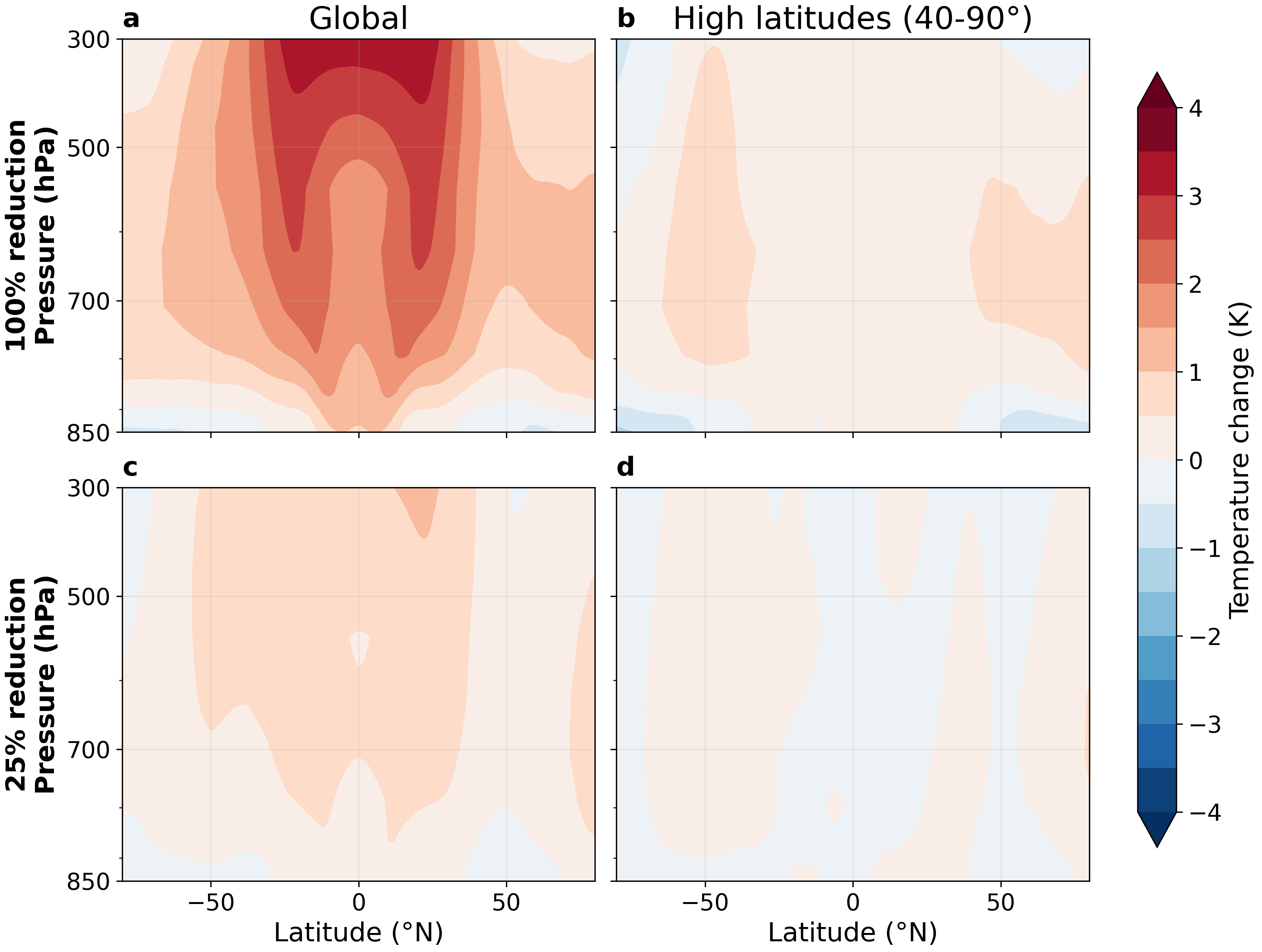}
    \caption{\textbf{Weaker and spatially restricted LW CRE perturbations reduce the atmospheric temperature response} Annual mean temperature anomaly (control-perturbed) in prescribed SST simulations with LW CRE reduced by 100\% (top) and 25\% (bottom), applied globally (left) or only at mid-to-high latitudes ($40^\circ$--$90^\circ$; right). The upper-left configuration closely resembles that of Li et al.~(2015) and reproduces a similar temperature response (their Fig.~5a).}
    \label{fig:ACRE}
\end{figure}

\begin{figure}
    \centering
    \includegraphics[width=\linewidth]{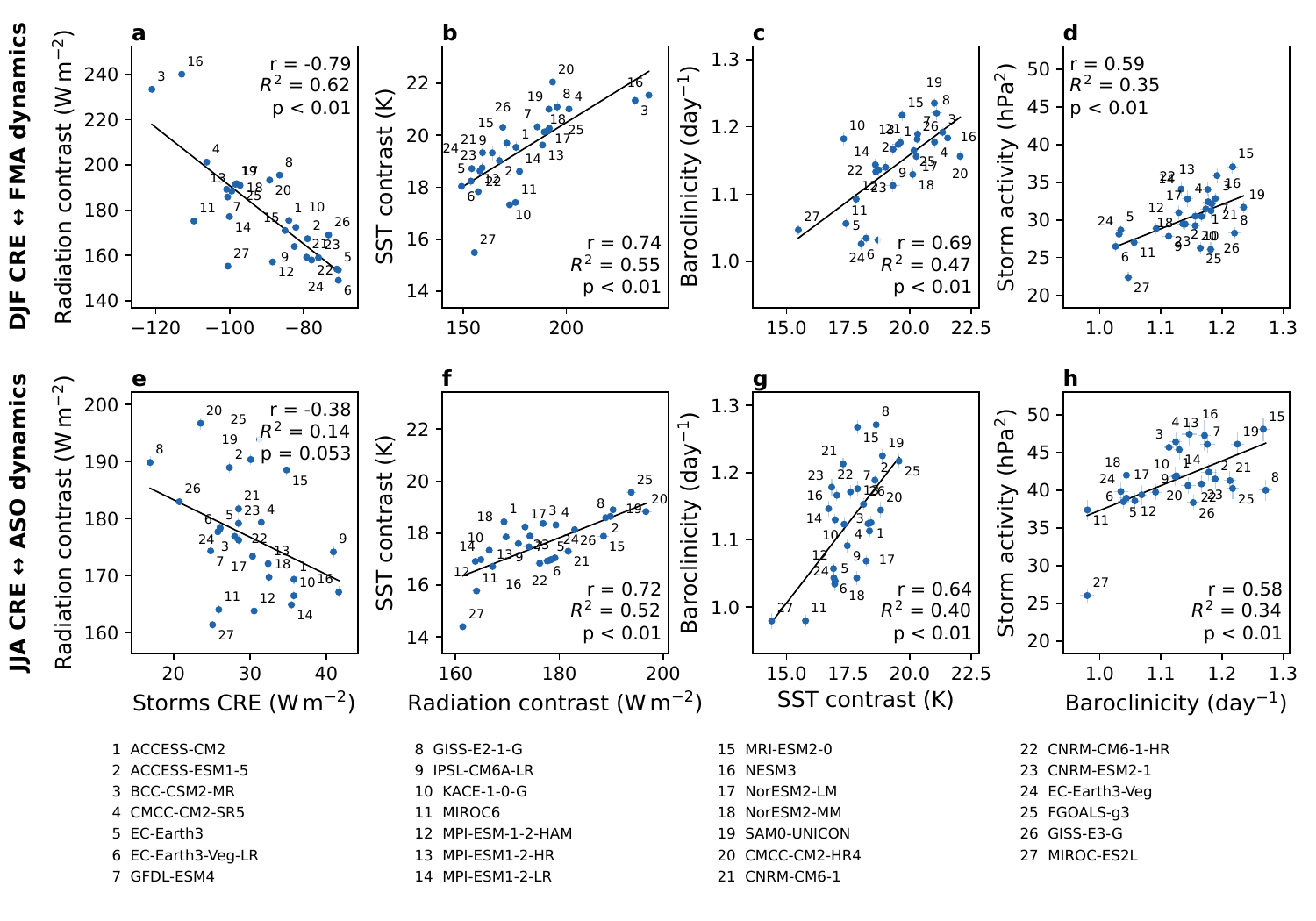}    
    \caption{\textbf{CMIP6 intermodel relationships along the proposed storm–cloud–ocean feedback pathway} (a) Relationship between storm-track CRE and the meridional surface radiative contrast during DJF. (b) Relationship between the DJF radiative contrast and the subsequent FMA sea surface temperature (SST) contrast. (c) Relationship between FMA SST contrast and atmospheric baroclinicity. (d) Relationship between FMA baroclinicity and storm activity. (e–h) Same as (a–d), but for winter radiative forcing (JJA) and late-winter and spring circulation (ASO).
    The analysis uses 27 CMIP6 historical simulations over 1985–2010, with one realization per model. 
    Storm-track CRE is defined as the net CRE (absorbed SW + LW) averaged over 50–60$^\circ$S. 
    Meridional radiative and SST contrasts are calculated as the differences between zonal-mean values at 30$^\circ$S and 60$^\circ$S, around peak storm-track CRE (Fig.~\ref{fig1}d). Atmospheric baroclinicity is calculated using Eq.~\ref{eq:baroc}, and storm activity is quantified by SLP variance. Black lines indicate ordinary least-squares linear fits, with Pearson correlation coefficients ($r$), the explained variance ($R^2$), and two-sided $p$-values. Error bars indicate 95\% confidence intervals estimated from interannual variability within each model.}
    \label{fig:CMIP6}
\end{figure}

\begin{figure}
    \centering
    \includegraphics[width=0.5\linewidth]{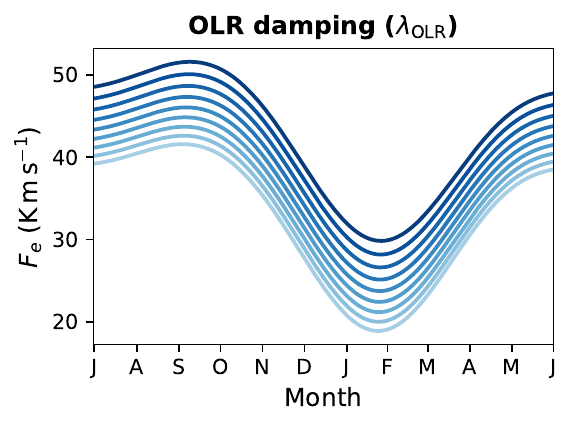}
    \caption{\textbf{Sensitivity of the conceptual model to OLR damping}
    Seasonal evolution of eddy heat transport $F_e$ for
     OLR damping coefficients $\lambda_{\mathrm{OLR}}$
    ranging from 0 to 4 in increments of 0.5 $~\mathrm{W\,m^{-2}\,K^{-1}}$ (dark to light blue).
    The OLR contribution to the temperature-contrast tendency is
    $-\lambda_{\mathrm{OLR}}\Delta T$.}
    \label{fig:OLR}
\end{figure}